\documentclass[aps,pre,twocolumn,groupedaddress,showpacs]{revtex4-1}

\usepackage[pdftex]{graphicx}
\usepackage{enumerate}

\usepackage{amssymb,amsfonts,amsmath}

\begin{document} 

\title{Fluid Flows Created by Swimming Bacteria Drive Self-Organization in Confined Suspensions}

\author{Enkeleida Lushi$^{1}$, Hugo Wioland$^{2}$, and Raymond E Goldstein$^{2}$}
\affiliation{$^{1}$School of Engineering, Brown University, 182 Hope Street, Providence, Rhode Island 02912, USA\\
$^{2}$Department of Applied Mathematics and Theoretical Physics, University of Cambridge, Wilberforce Road, Cambridge CB3 0WA, United Kingdom}

\begin{abstract}
Concentrated suspensions of swimming microorganisms and other forms of active matter are known to display complex, 
self-organized spatio-temporal patterns on scales large compared to those of the individual motile units.   Despite 
intensive experimental and theoretical study, it has remained unclear the extent to which the hydrodynamic flows
generated by swimming cells, rather than purely steric interactions between them, drive the self-organization.  
Here we utilize the
recent discovery of a spiral-vortex state  in confined suspensions of \textit{B. subtilis} to study this issue in detail.   
Those experiments
showed that if the radius of confinement in a thin cylindrical chamber is below a critical value the suspension will
spontaneously form a steady single-vortex state encircled by a counter-rotating cell boundary layer, with 
spiral cell orientation within the vortex.   Left unclear, however, was the flagellar orientation, and hence the 
cell swimming direction, within the spiral vortex.  
Here, using a fast simulation method that captures oriented cell-cell 
and cell-fluid interactions in a minimal model of discrete-particle systems, we predict the striking, counterintuitive result 
that in the presence of 
collectively-generated fluid motion the cells within the spiral vortex actually swim upstream against those flows.
This is then confirmed by new experiments reported here, which include measurements of flagella bundle orientation and
cell tracking in the self-organized state.  These results highlight the complex interplay between cell orientation and
hydrodynamic flows in concentrated suspensions of microorganisms.
\end{abstract}


\maketitle

\noindent{\it Significance statement: 
The collective dynamics of swimming microorganisms exhibits a complex interplay with the surrounding fluid: the motile cells 
stir the fluid, which in turn can reorient and advect them. This feedback loop can result in long range interactions between the 
cells, an effect whose significance remain controversial. We present a computational model that takes into account these 
cell-fluid interactions as well as cell-cell forces, and which predicts counterintuitive cellular order driven by 
long-range flows. This is confirmed with new experimental studies which track the orientation of cells in a confined, dense 
bacterial suspension.}

\vskip 0.5cm

In the wide variety of systems termed `active matter' \cite{Ramaswamy10,Marchetti13}
one finds the spontaneous appearance of coherent dynamic structures on scales large compared to the individual motile units.
Examples range from polar gels \cite{Kruse04, Furthauer12}, bacterial suspensions
\cite{Dombrowski04,Tuval05,Sokolov07,Cisneros07,Swinney10,Sokolov12} 
and microtubule bundles \cite{Sanchez12} to cytoplasmic streaming \cite{Woodhouse12, Woodhouse13}. 
At high concentrations, suspensions of rod-like bacteria are known to arrange at the cellular scale 
with parallel alignment as in nematic liquid crystals \cite{Dombrowski04,Volfson08}, but 
with local order that is polar, driven by motility \cite{Ginelli10, WensinkEtAl12}. 
At meso- and macroscopic scales, coherent structures such as swirls, 
jets, and vortices at scales $10\,\mu$m to $1\,$mm have been experimentally observed
\cite{Dombrowski04,Tuval05,Sokolov07,Cisneros07,Swinney10,Sokolov12}.
Many studies have focused on how complex cell interactions can give rise to macroscopic organization and ordering, and 
the role of self-generated fluid flows in the dynamics of dense suspensions is still under debate 
\cite{SimhaRama02,Cisneros07,Sokolov12, Fielding12, Dunkel13, Aranson13, Zottl14}. 
This is due in part to the inherent complexity of the systems under investigation, and the difficulty of making faithful
mathematical models.

Microswimmers such as {\it E. coli}, {\it B. subtilis} or {\it C. rheinhardtii} produce dipolar fluid 
flows through the combined action of their flagella and cell body on the fluid.  In the far field, they are well-described as 
`pusher' or `puller' 
stresslets \cite{Drescher10,Guasto10,Drescher11},
corresponding to the case of flagella behind or in front of the cell body. These fluid flows 
affect passive tracers \cite{Leptos2009,Rushkin} as well as swimmers: their motion is subject to convection and shear reorientation 
induced by neighboring organisms, which can lead to complex collective organization. Macroscopic fluid flows emerge from the collective 
motion of a colony of motile bacteria and the suspension can exhibit a quasi-turbulent dynamics \cite{Dombrowski04}. 
Microorganisms like {\it B. subtilis} live in porous environments, such as soil, where contact with surfaces is inevitable as 
mesoscale obstacles and confinement are the norm. Recent experiments give insight into the interactions of single microorganisms 
with surfaces \cite{DiLuzio05, Li08,Drescher11,Kantsler13}, yet suspension dynamics in confinement has only begun to be investigated 
\cite{WiolandEtAl2013} and the role of the collectively-generated fluid flows in the macroscopic organization has yet to be fully understood.

\begin{figure*}
\centering
\includegraphics[width=1.9\columnwidth]{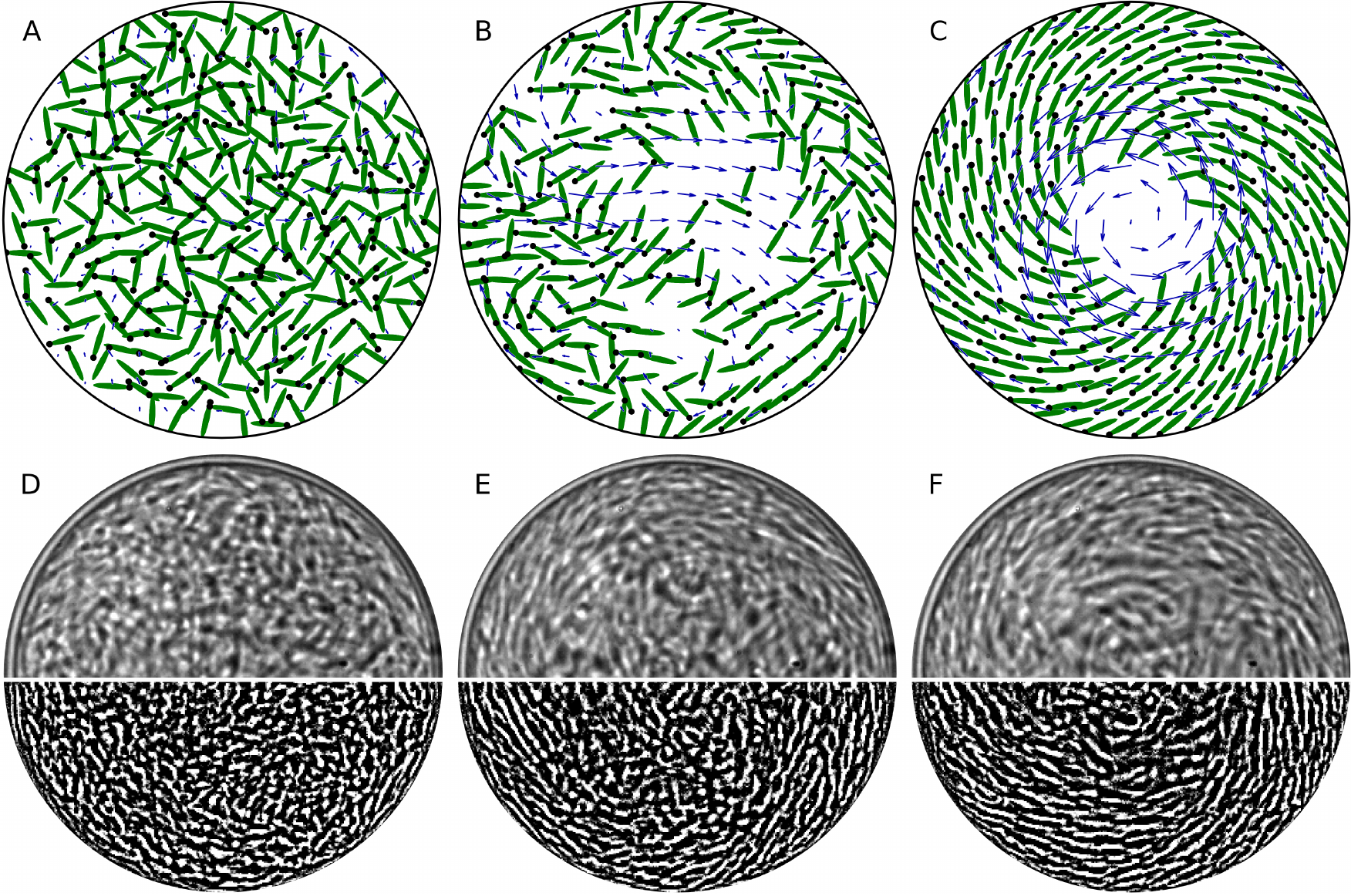}
\caption{Snapshots of the bacterial suspension self-organization from simulations (A-C) and experiments (D-F). A-C: an 
initially isotropic suspension of micro-swimmers inside a circle with diameter $12\ell$ ($\ell$=individual swimmer length). Black 
dots indicate the swimming direction. The swimmer-generated fluid flow is shown superimposed in each plot (blue arrows). D-F: a 
dense suspension of \textit{B. subtilis} in a drop, $70\,\mu$m in diameter. Top halves: bright field, 
bottom halves: images processed by edge-detection filtering. Initial disordered state is obtained by shining a blue 
laser that causes cells to tumble. In both simulations and experiments the suspension organization initiates at 
the boundary, as seen in (B,E). See also SI Movie 1.}
\label{Fig1}
\end{figure*}

Recently, Wioland {\it et al.} \cite{WiolandEtAl2013} showed that a dense suspension of {\it B. subtilis}, 
confined into a flattened drop, can self-organize into a spiral vortex, in which a boundary layer of cells at the drop 
edge moves in the opposite direction to the bulk circulation.
This spatio-temporal organization is driven by the presence of the circular boundary and the interactions
of bacteria with it.
At the interface the packed cells move at an angle to the tangential that is 
dictated by the drop curvature, swimmer size and shape. This macroscopic nonequilibrium pattern and double-circulation were not 
anticipated by theory and have not been seen in any simulations of discrete particle systems.
This is due to the computational difficulty of capturing both confinement and complex interactions between 
elongated swimmers.  Although previous simulations have demonstrated the importance of hydrodynamics in populations of spherical 
squirmers \cite{IshikawaPedley08} and rod-shaped swimmers \cite{SaintillanShelley1}, they do not consider boundary 
effects and the elongated shape of the swimmers in the steric interactions.
On the other hand, continuum models of motile suspensions that include fluid dynamics, and have been successful in explaining 
large-scale patterns \cite{SaintillanShelley1,SaintillanShelley2}, have either ignored confinement or interactions with 
surfaces, or, if addressing confinement \cite{Ravnik13}, have imposed boundary conditions that generally do not resolve 
the orientations of the bacteria at the interface. 
Thus, the conditions at boundaries and microscopic interactions between cells warrant careful consideration 
in the modelling of these suspensions so that the macroscopic dynamics and organization are correctly captured.

Here, we elucidate the origin and nature of the spontaneous emergence of the spiral vortex and cellular organization in a 
confined motile suspension. A computational model is described for bacterial suspensions in which the 
direct and hydrodynamic interactions between the swimmers and the confining circular interface can be tuned. The cells are 
represented as oriented circles or ellipses subject to cell-cell and cell-fluid interaction, while the fluid flow is the 
total of the pusher dipolar fluid flows produced by each swimmer's locomotion. It is shown that while some circulation 
under conditions of confinement may arise with direct interactions only, hydrodynamics are necessary and crucial to reproduce 
and explain the double circulation that is observed experimentally. Simulations (Fig. 1A-C) are able to reproduce the emergence 
of the spiral vortex from an isotropic state (Fig. 1D-F), and give insights into the origin of the microscopic organisation 
of the bacteria in the drop. The computational results show the remarkable feature that cells in the bulk of the drop swim 
{\it against} the stronger colony-generated fluid flow and thus have a net {\it backward} motion.
We confirm this observation by measuring the orientation of the cells and of their flagella through suitable
fluorescent labelling methods.
 
\section{Model and Simulations}  

Since no vertical motion of the swimmers was observed experimentally  \cite{WiolandEtAl2013} we assume the dynamics is 
confined to a plane and the simulations are therefore implemented in two-dimensional domains to increase the computational 
speed. We consider $M$ pusher swimmers 
immersed in a fluid contained inside a circle. 
Each swimmer is modelled as a slender ellipse with length $\ell=1$, width 
$w=\ell/6$ (or $w=\ell$ for disks) that propels itself along its main axis. The dynamics of each is expressed 
in terms of its center of 
mass position $\mathbf{X}_i$ and orientation $\mathbf{P}_i$ \cite{LP2013},
\begin{align}
\partial_t \mathbf{X}_i &= V\mathbf{P}_i + \mathbf{u} +  \Xi^{-1}_i\sum_{j \neq i} \mathbf{F}^e_{ij}\label{xdot}~, \\
\partial_t \mathbf{P}_i &= (\mathbf{I} - \mathbf{P}_i\mathbf{P}_i^T)(\gamma \mathbf{E} + \mathbf{W}) \mathbf{P}_i + k\sum_{j \neq i} 
\mathbf{T}^e_{ij} \times \mathbf{P}_i.\label{pdot}
\end{align}
Eq. (\ref{xdot}) describes self-propulsion with constant speed (chosen as $V=1$ without loss of generality) 
along the cell direction $\mathbf{P}_i$, advection 
by the fluid velocity $\mathbf{u}$, and pairwise repulsion with force $\mathbf{F}^e_{ij}$ between swimmers. Here, 
$\Xi = m_{||}\mathbf{P}_i\mathbf{P}_i^T + m_{\perp}(\mathbf{I} - \mathbf{P}_i\mathbf{P}_i^T)$ with $m_{\perp}$=$2 m_{||}=2$ for 
elongated ellipses and $m_{\perp}$=$m_{||}=1$ for disks \cite{KimKarrila}. The first term of Eq. (\ref{pdot}) 
describes rotation of the 
particle by the fluid flow $\mathbf{u}$ with 
$2\mathbf{E}=\nabla \mathbf{u}+\nabla \mathbf{u}^T$,  $2\mathbf{W}=\nabla \mathbf{u}-\nabla \mathbf{u}^T$; $\gamma\approx 0.95$ for 
ellipses with aspect ratio 6 and $\gamma=0$ for disks. The last term of Eq. (\ref{pdot}) describes swimmer rotations due to torques 
from direct interactions with neighbors; $k$=$6$ for elongated ellipses and $k=0$ for disks. The purely repulsive steric forces 
$\mathbf{F}^e_{ij}$ and torques $\mathbf{T}^e_{ij}$ are obtained using methods described elsewhere \cite{ConstanzoEtAl2012}. Each swimmer 
is discretized into $n_b$ beads ($n_b=6$ for ellipses, $n_b=1$ for disks). Beads from different swimmers interact by with a soft 
capped Lennard-Jones potential; this allows some overlaps. Noise terms are not included. 

The swimmer-driven fluid velocity $\mathbf{u}$ is governed by the (non-dimensional) 2D Stokes Equations with an extra stress,
\begin{align}\label{stokes}
-\nabla^2 \mathbf{u} + \nabla q = \nabla \cdot \sum_{i} \mathbf{S}^a_i \delta (\mathbf{x}-\mathbf{X}_i), \quad
\nabla \cdot \mathbf{u} = 0.
 \end{align}
Here $q$ is the fluid pressure to account for the fluid incompressibility and $\mathbf{S}^a_i$=$\alpha \mathbf{P}\mathbf{P}_i^T$ 
denotes the active stress tensor resulting from the swimmer locomotion in 2D with non-dimensional stresslet strength $\alpha\approx -1$ 
for a pusher swimmer with length $\ell=1$ and speed $V=1$. Eqs. (\ref{xdot}-\ref{pdot}) are integrated in time and 
the instantaneous fluid flow that swimmers collectively generate is obtained by solving Eq. (\ref{stokes}) on an underlying uniform 
Eulerian grid \cite{LP2013}. The interpolation of the fluid velocity $\mathbf{u}$ to the swimmers' positions $\mathbf{X}_i$ and 
the extrapolation of the active stresses $\mathbf{S}^a_i$ onto the Eulerian grid $\mathbf{x}$ are done using an Immersed Boundary 
method framework \cite{Peskin2002} with a discretized delta function $\delta (\mathbf{x}-\mathbf{X}_i)$. 
Essentially, the  fluid velocity $\mathbf{u}$ is the superposition of the pusher-like dipolar flows generated by each swimmer.

We use the method of images for swimmers at the drop boundary \cite{DiLeonardo2011}, which in experiments is an oil-water interface.
In a circle of radius $R$, at each time-step, if a swimmer $i$ is within $3\ell$ of the surface, 
then an approximate mirror swimmer is placed outside the circle at $R^2/||\mathbf{X}_i-\mathbf{X}_{center}||$, 
with mirror orientation $\mathbf{P}_i - 2(\mathbf{X}_i-\mathbf{X}_{center})/||\mathbf{X}_i-\mathbf{X}_{center}||$. The steric forces, 
torques and the fluid velocity $\mathbf{u}$ are calculated for the swimmers and their images, 
while Eqs. (\ref{xdot}-\ref{pdot}) are integrated in time only for the actual swimmers. This approach approximates simultaneously 
and at low computational cost an effective confinement as well as no-stress condition at the drop boundary.  
This boundary condition is appropriate, as we observed that the surrounding oil in experiments was set in motion
by the drops of bacterial suspension.

\section{Simulation Results} 

We first describe computational results in an unconfined periodic domain. With neglect of hydrodynamics ($\alpha=0=\mathbf{u}$), 
the suspension exhibits swarming at low concentration (Fig. 2A and SI Movie 1) or a stable `bionematic' state \cite{Dombrowski04} 
at higher concentration, as classified by \cite{WensinkEtAl12} and seen in swarming colonies on surfaces \cite{Henrichsen1972}. 
Introducing hydrodynamics (Fig. 2B and SI Movie 1) destabilizes these two states to generate a turbulent dynamics qualitatively similar 
to experimental observation \cite{Dombrowski04, Cisneros07}. Remarkably, hydrodynamics disrupts bacterial clusters, as 
also suggested by a squirmer model \cite{Fielding12}. This disruption of long-range polar order is a consequence 
of an instability that has been previously analyzed \cite{SimhaRama02, SaintillanShelley1,SaintillanShelley2}.

Next, consider the case of confined suspensions.
Without fluid interactions ($\alpha=0=\mathbf{u}$), Fig. 2C (see also SI Movie 3) shows cells concentrating or jamming 
at the drop boundary with small unidirectional circulation. With more realistic conditions (ellipses, direct and hydrodynamic 
interactions) we observe in Fig. 2D (and SI Movie 4) a spiral vortex similar in form and dynamics to that of experiments (Fig. 2H).
For disk-shaped pusher swimmers ($\gamma=0$, $\alpha=-1$, $\mathbf{T}^e_{ij}$=$\mathbf{0}$) subject to reorientation and advection 
by the fluid flows they create, Fig. 2E (and SI Movie 5) shows that while the alignment between neighbouring cells is lost, 
swimmers form unstable layers with very small transient circulation.
These three configurations show that while steric interactions force local alignment of ellipsoidal swimmers, 
it is the collectively-generated fluid flow that produces the large-scale organization and double-circulation. 
In fact, in simulations where alignment with the flow but not through steric interaction was considered 
(i.e. setting $\gamma \approx 1$ but $k=0$, $\mathbf{T}^e_{ij}$=$\mathbf{0}$) the spiral organization and double-circulation 
are still obtained (Fig. 2F and SI Movie 6, cells are spaced further apart due to isotropic steric repulsions with a larger radius).
These model parameters would be appropriate to the description of spherical bacteria, whose collective behavior has recently been 
studied in the absence of confinement \cite{Beer}. In Fig. 2G (and SI Movie 7) we show that a dilute suspension of ellipse-shaped 
swimmers also orders into a spiral vortex and self-generates bidirectional fluid flows. As an aside, we note that 
circulation has been observed in confined systems of self-propelling disks with prescribed alignment interactions and 
possibly noise terms \cite{Grossman08}, but that circulation is unidirectional, as in Fig. 2C.

\begin{figure}
\centering
\includegraphics[width=0.45\textwidth]{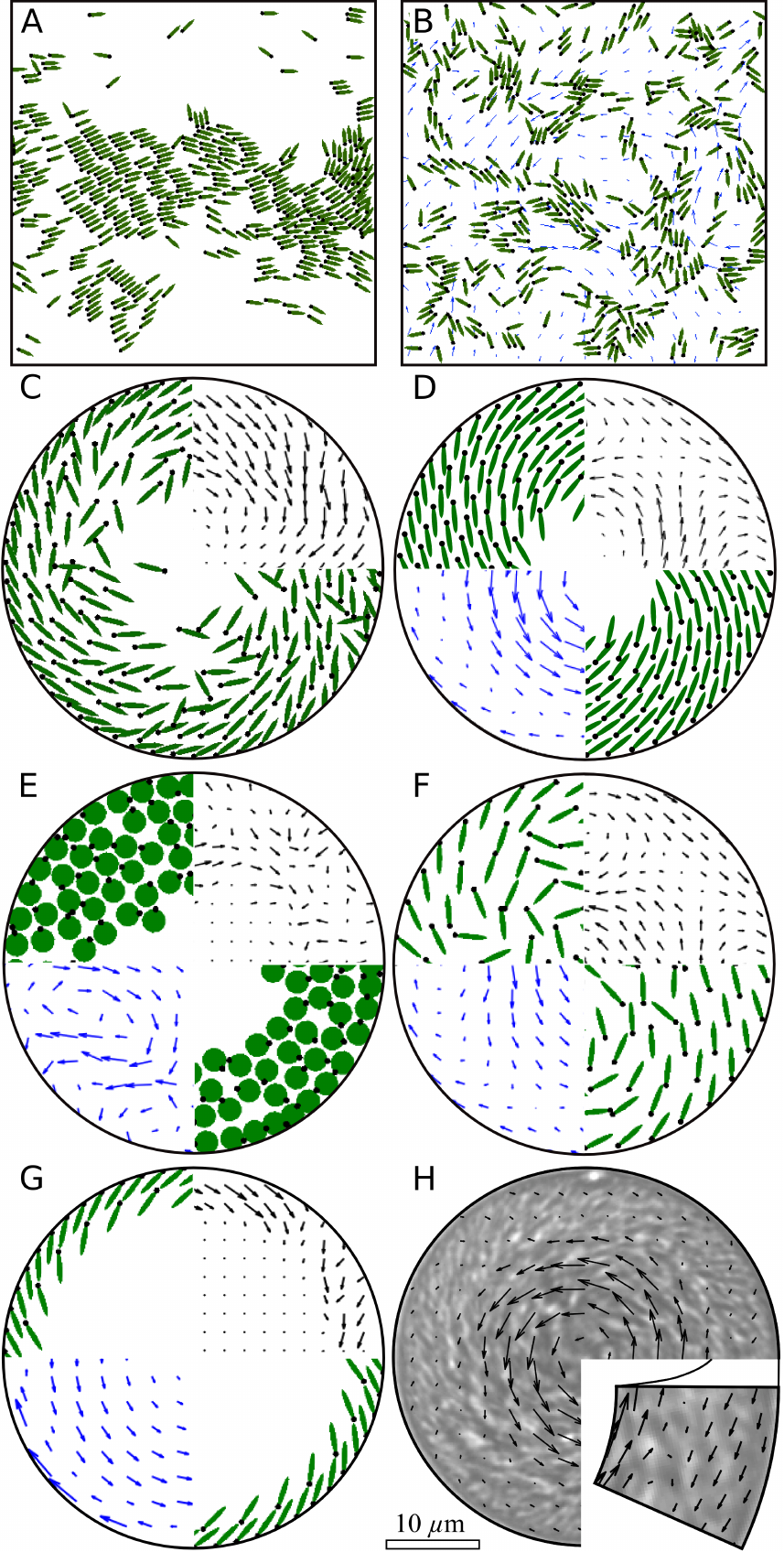}
\caption{Suspension organization in periodic domains and inside circular confinement. (A) Self-propelled ellipses interacting without 
hydrodynamics achieve a swarming or bionematic state when in a periodic domain. (B) Pusher swimmers in a periodic domain exhibit a turbulent 
dynamics and are less clustered. (C) Self-propelled ellipses interacting sterically without any hydrodynamics. (D) Self-propelled ellipses 
interacting sterically with hydrodynamics. (E) Circular pusher swimmers. (F) Pusher swimmers with isotropic steric repulsions but alignment with fluid flow ($\gamma \approx 1, k=0, \mathbf{T} = \mathbf{0}$). 
(G) Dilute suspension of ellipsoid pusher swimmers. (H) Bacterial flow measured in experiments by PIV \cite{WiolandEtAl2013}. 
The upper-right insets in (C-G) indicate the swimmer net circulation direction and plot the mean swimmer motion with arrows magnified by 4,1,13,1,2 respectively. Lower-left inset in (D-G) shows the fluid flow velocity arrows magnified by 1,5,1,2 respectively. 
See SI Movies $1-7$.}
\label{Fig2}
\end{figure}

\section{Emergence of Organization}

Bacteria move by swimming and through advection by the local fluid flow; the balance between the two yields the observed direction 
of motion. In drops of bacterial suspension, particle image velocimetry (PIV) measurements reported previously \cite{WiolandEtAl2013} 
showed that a boundary layer circulates in a direction opposite to that of the bulk but it was not possible to resolve whether 
swimming or advection dominates the overall motion, and in particular in which direction cells point.

To understand how the spiral order and double circulation arise, we consider suspensions of increasing density. 
When a few cells are trapped in a drop, they swim to the oil interface and slide along it at a small angle, as in Fig. 2G. With 
more cells added, 
they form clusters sliding along the boundary (akin to those seen with self-propelling rods in channels \cite{Wensink08}). 
The clusters finally merge to form the circulating outer boundary layer, as seen in Fig 2G. (In the images of Fig. 1B\&E we note that this layer 
is the first to form.) Bacteria point outward with an angle characteristic of the spiral pattern. As pusher swimmers, they push fluid 
backwards and the added effect produces the drop bulk fluid flow that is in the opposite direction to the swimmer circulation.

Upon increasing the concentration to the dense regime, additional cells arrange into more layers with an angle dictated by 
steric interactions, thus reproducing the spiral arrangement. While bulk cells were first proposed to be pointing inward 
\cite{WiolandEtAl2013}, simulations show instead that almost all cells point {\it outward} and swim in the same direction 
(clockwise in Fig. 2D). Moreover, the fluid flow is in the opposite direction to the bulk swimmers' orientation and - in the inner part of the drop - 
is strong enough to counterbalance the swimming speed. This is illustrated in Fig. 1 where central cells point clockwise but overall move 
counterclockwise. While the macroscopic suspension dynamics is in agreement with experiments (Fig. 2H), new experiments are required 
to determine the actual cell orientation and to test the predicted arrangement. 

\begin{figure*}
\centering
\includegraphics[width=1.7\columnwidth]{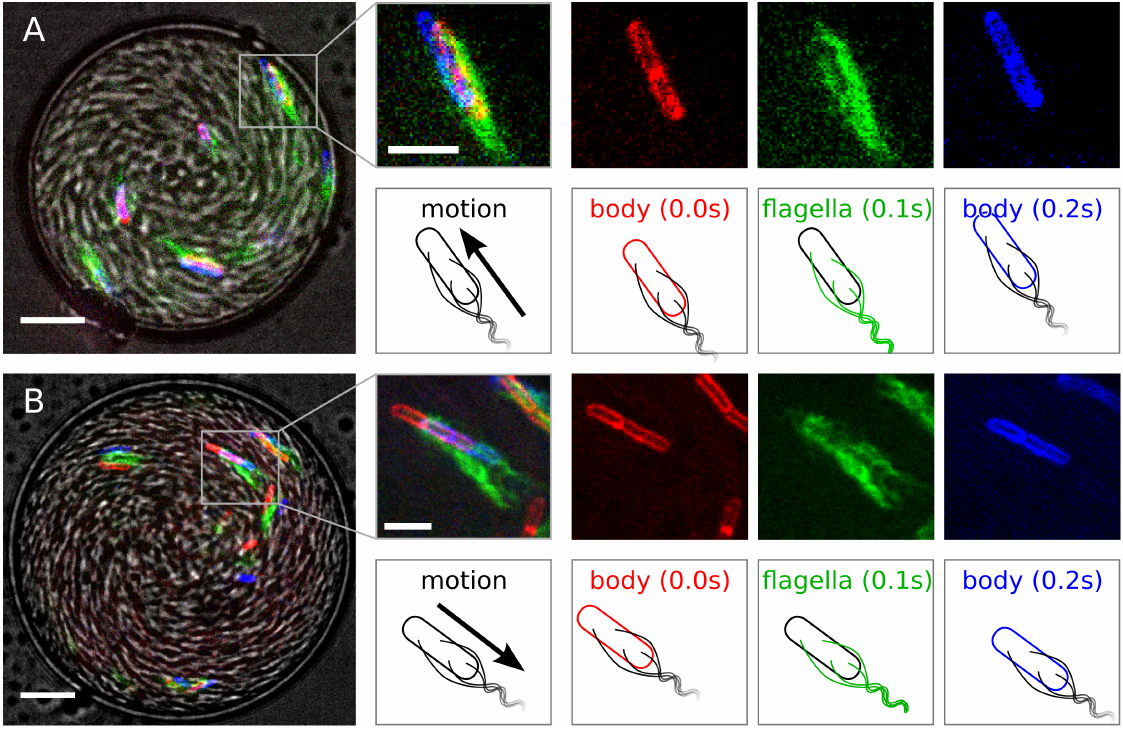}
\caption{Drop overview in gray scale: bright field image. Positions of the membrane (false coloured red at time $t$=$0$, blue at $t$=0.2$s$) 
and flagella (false coloured green at $t=0.1s$) dyes help determine the cell orientation. (A) Forward motion: cell at the oil 
interface both point and move to the top left corner. (B) Backward motion: the cell is pointing to the top left corner while moving 
overall in the opposite direction.  Bar at the drop-images: 10$\mu$m. Bar at the individual bacterium images: 5$\mu$m. }
\label{fluo}
\end{figure*}

\section{Experiments}  
To determine precisely the cell configuration and orientation in the spiral vortex state, we used a mutant of 
{\it B. subtilis} (DS1919 3610, a generous gift of H.C. Berg \cite{Blair2008}), labelled with Alexa Fluor 488 C5 maleimide 
on the flagella and FM4-64 on the cell membrane, following the protocol of {Guttenplan et al.} 
\cite{Guttenplan2013}. From these two-colored bacteria, mixed with a large amount of wild-type cells (strain 168), we 
form numerous drops $10-100\,\mu$m in diameter and $\approx\! 25\,\mu$m in height in an emulsion, the background liquid of
which is mineral oil \cite{WiolandEtAl2013}. A sequence of four images was taken: (i) in bright field to determine to spatial 
organization (grey scale in Figs. 3A\&B), (ii) of the membrane (FM4-64, false colored red), (iii) of the 
flagella (Alexa-488, false colored green), and (iv) again of the membrane (FM4-64, false colored blue). From these we determine 
the cell position, overall motion and swimming direction.

Fig. 3A highlights a cell at the oil interface. Both the cell motion and swimming direction are toward the top-left corner.  
Fig. 3B highlights a bacterium in the bulk. The two images of the membrane indicate that the cell is moving toward the lower right 
corner. Yet, the relation between the flagella position and the mean membrane position and the flagella bundling at the rear of the 
bacteria reveal that the cell is pointing to the top-left corner, in a direction opposite to its motion.   
These results, consistently found by sampling over $20$ cells, confirm the prediction from simulations: while all the bacteria point 
in the same direction (outward), the bulk micro-swimmers move overall in a backward fashion, opposite to the boundary layer motion.

\section{Quantification of Spatial Order}
As shown in previous sections, experiments and simulations with both steric and hydrodynamical interactions are in qualitative agreement 
on both micro- and macroscopic scales.  In this section we consider quantitative measures of the spatial order in the numerical studies
and compare them to the experimental results reported previously \cite{WiolandEtAl2013}.
In experiments drops show stable circulation when the confining chamber is $30-70\,\mu$m in diameter. To quantify the order in simulations 
we introduce the 
vortex order parameter
\begin{equation}\label{phi}
\Phi = \frac{1}{1-2/\pi}\left(\frac{\Sigma_i |\mathbf{v}_i \cdot \mathbf{t}_i|}{\Sigma_j || \mathbf{v}_j||} - \frac{2}{\pi} \right)
\end{equation}
with $\mathbf{v}_i$ the bacterial overall motion and $\mathbf{t}_i$ the azimuthal unit vector. $\Phi=1$ for purely azimuthal flows, $\Phi=0$ for 
disordered chaotic flows and $\Phi<$0 for mostly radial flows. $\Phi$ is computed for drops with diameters between $4\ell\approx 20\,\mu$m and 
$25\ell\approx 125\,\mu$m for dense (area fraction $\approx$25$\%$) and semi-dilute suspensions (area fraction $\approx 10\%$).

Fig. 4 shows a first transition from random to vortex state around $d=7\ell$. For dense suspensions the plot reveals that a 
highly-ordered single-vortex state with $\Phi>0.7$ is achieved in drops with diameter $d=7-16\ell$ (versus $30-70\,\mu$m, 
$\approx 6-14\ell$ in experiments). In experiments and simulations, we observe that turbulence arises in the center of the 
largest drops. In the case of dilute or semi-dilute suspensions (as in Fig. 2G), this center is depleted in cells, thus 
leading to ordered states even for $d>14\ell$.
 
\begin{figure}[t]
\centering
\includegraphics[width=0.45\textwidth]{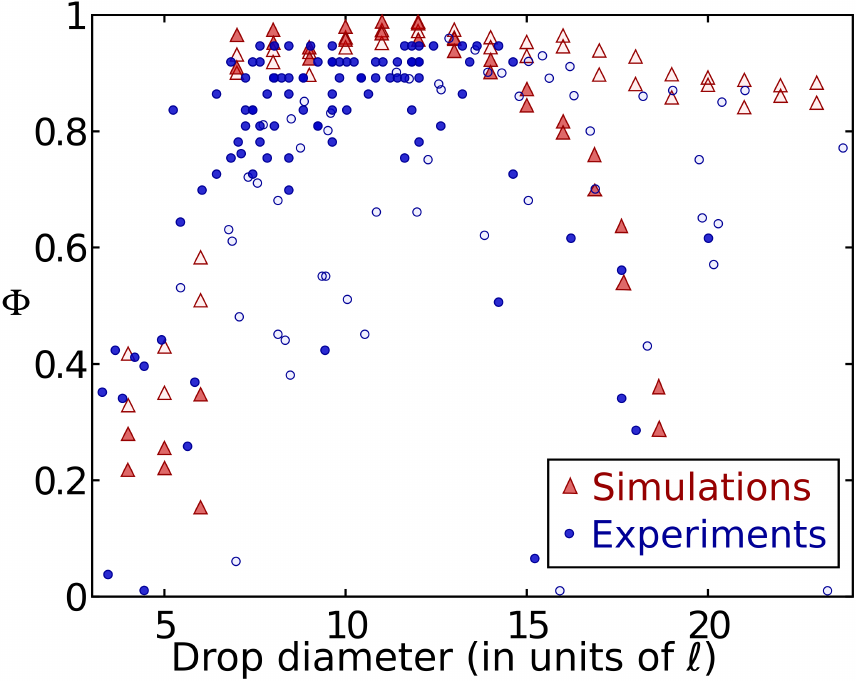}
\caption{A stable vortex forms for a range of drop diameters. The vortex order parameter $\Phi$  for dense (solid color) or dilute (faded color) 
suspensions in drops with various diameters for both simulations (triangles) and experiments (circles).}
\label{order}
\end{figure}

As seen in Fig. 1, in both simulations and experiments the cell orientation is not parallel to the direction of the fluid 
circulation. We examine the azimuthally-averaged swimmer orientation angle $\theta_m$ for the layer of cells at the boundary. Not surprisingly, 
the effect of the surface curvature makes this angle higher for smaller drops and smaller for larger drops. In experiments it ranges 
from $\theta_m\simeq 35^o$ for drop diameters $d=30\,\mu$m to $\theta_m\simeq 10^o$ for 
$d=70\,\mu$m \cite{WiolandEtAl2013}. In simulations it ranges from $\theta_m\simeq 42^o$ for circle drops with diameter 
$d=7\ell$ to $\theta_m\simeq 36^o$ for $d=16\ell$. Though the boundary behavior and swimmer angles are qualitatively similar to the experiments, 
they do not match quantitatively due to simplifications in the model. Including short-range hydrodynamics between the cells would likely
remedy this discrepancy.

\section{Discussion} 

We have presented a minimal model and simulation method for
micro-swimmer suspensions that includes direct cell-cell, cell-fluid
interactions and swimmer-generated flows. The method, though minimal
and in two-dimensions, captures well the dynamics seen in experiments
on confined bacterial suspensions. 
In agreement with previous simulations, we show here that the long-range hydrodynamic 
interactions are crucial to reproduce the organization and circulation that are observed in experiments.
In periodic domains the swarming
states predicted by active matter theories are disrupted by
hydrodynamics, resulting in a more turbulent suspension behavior.
Under circular confinement, although direct cell interactions lead to
local cell alignment, large scale order appears only when the swimmer
motion is coupled to the fluid dynamics. Simulation results not only
agree qualitatively with the experiments, they also highlight the
microscopic bacterial organization. In particular, the cells in the
bulk are shown to swim against the collectively-generated fluid flow,
a result that was not foreseen in previous publications.
We confirmed that prediction by experimentally recording both motion and swimming directions, which to our knowledge has never been 
done in dense bacterial suspensions. To do so we tagged and tracked the relative positions of the cells body and flagella.

These results emphasize the necessity to include more realistic
hydrodynamic interactions in active fluid particle simulations and 
also continuum theories \cite{SimhaRama02, SaintillanShelley2}. Ultimately, a
closer comparison to experiments requires 3D particle simulations, more accurate
descriptions of the fluid flows generated by the swimmers in the bulk
and near boundaries, and possibly an accounting of the the swimmers'
geometry and flagella.
This model could be adapted to a 3D domain, changing the packing of
the cells but also the fluid flow they generate: in 2D domains a
swimmer dipolar flow decays with distance from the cell as $1/r$ 
instead of the $1/r^2$ decay in 3D which is more appropriate for the experimental situation.
The present model does not account for close-range and lubrication
hydrodynamics or Brownian noise effects. Recent simulations of
spherical squirmers at high packing fractions \cite{Fielding12,
Zottl14} show that including close-range hydrodynamics significantly
affects the suspension behavior.
Although simplified, the model described here has been shown to give good insights into
the dynamics of micro-swimmer suspensions and
could be applied to more complex geometries to study
microscopic interactions and ordering that are difficult to visualize
experimentally.

\section{Materials and Methods}
\subsection{Experimental Protocol}
We use two {\it B. subtilis} strains: the wild-type 168 strain and the mutant amyE::hag(T204C) DS1919 3610 (generous gift of H. Berg), 
both grown in standard Terrific Broth (TB, Sigma) at 35$^{\circ}\,$C on a shaker. An overnight culture was diluted 200$\times$ and 
grown for 5$\,$h until the end of exponential growth when the proportion of motile cells is maximal \cite{kearns2005}. 

To label mutant bacteria, we followed the protocol of Guttenplan {\it et al.} \cite{Guttenplan2013}. One millilitre of the suspension 
was centrifuged (1000g, 2 min) and resuspended in 50 $\mu$L of PBS containing $5\,\mu$g/mL Alexa Fluor 488 C5 Maleimide (Molecular Probes) 
and incubated at room temperature for 5 min, to stain the flagella. Bacteria were then washed in 1mL PBS and resuspended in PBS 
containing 5 $\mu$g/mL FM4-64 (Molecular Probes) for membrane staining.

A dense suspension of bacteria was prepared by centrifuging 10 mL of wild-type {\it B. subtilis} (1500g for 10 min). If necessary, 
a small volume of stained bacteria was added to the pellet, which was then mixed into 4 volumes of mineral oil containing 10 mg/mL 
diphytanoyl phosphatidylcholine (DiPhyPC, Avanti) to prevent drops from coalescing. The emulsion was then created by gently 
pipetting the suspension and placing it between two silane-coated coverslips, creating numerous flattened drops, $\approx$ 25 $\mu$m in height and 
10-100 $\mu$m in diameter.

Bright field movies were acquired at 125 fps with a high-speed camera (Fastcam, Photron) on an inverted microscope 
(Cell Observer, Zeiss), using a 100$\times$ oil-immersion objective and analyzed with Matlab mPIV algorithm \cite{mori2006}.
To observe the emergence of order (Fig. 1. and SI Movies 1 and 2), we shined blue light on the drop for a few seconds. 
Bacteria naturally react by tumbling \cite{Macnab1974}, thus disorganizing the drop.

To measure the swimming and motion directions of the cells, we imaged stained mutant \textit{B. subtilis} on a confocal spinning disc 
microscope. To increase the resolution, images were taken at the bottom of the drop. We excited both fluorophores with a 488 nm laser and 
filtered the emission with a GFP filter cube (barrier filter 500-550 nm, Zeiss) for Alexa Fluor 488 C5 Maleimide and DsRed filter cube 
(barrier filter 570-640 nm, Zeiss) for FM4-64. Images were taken every 0.1s (limited by the acquisition rate), filtering first for the 
membrane (false colored red), flagella (false colored green) and again membrane (false colored blue, Fig. 3). The direction of 
bacterial motion was determined from the membrane displacement and the swimming direction from the relative position of the flagella to the 
average membrane position.

\subsection{Simulations} To calculate the repulsive forces and torques between neighbouring swimmers, we use the method of Constanzo {\it et. al.} 
\cite{ConstanzoEtAl2012} and discretize each swimmer into $n_b$ beads of diameter $\ell/n_b$. Beads of different swimmers interact with each 
other via a capped Lennard-Jones potential,
\begin{align}
\Psi_{LJ}^{\alpha}(r) = \begin{cases} 8 \epsilon \left( \frac{ 2r_c^{12}}{(r^2+\alpha^2)^6}  -\frac{ r_c^{6}}{(r^2+\alpha^2)^3}
\right) &\mbox{if } r \leq r_{c} \\
0 & \mbox{if } r > r_{c}. \end{cases} 
\end{align}
where $r$ is the distance between the bead centers, $r_c=2r_b$ is the cut-off distance, $r_b$ is the bead radius and 
$\alpha=r_c[(1/2)^{1/3}-1]^{1/2}$ is the capping or smoothing factor. The smoothing of the potential allows for larger time-steps when 
integrating Eqs. (\ref{xdot}-\ref{pdot}), but it comes at the expense of the swimmers overlapping or possibly escaping the confinement. 
The effective bead radius is then $\approx r_c/2$, giving the swimmer an effective thickness of $r_c$ and effective aspect ratio of $\ell/r_c$.





\vskip 0.5cm

\begin{acknowledgments}
We thank Howard Berg, Linda Turner and Daniel Kearn for generously providing us with the mutant bacterium strain amyE::hag(T204C) 
and help on the staining protocol. We thank I. Aranson, J. Dunkel, M. Shelley for helpful discussions. E.L. acknowledges the support 
of a David Crighton fellowship and the hospitality of DAMTP at the University of Cambridge. H.W. and R.E.G. were supported in part by 
the EPSRC and ERC Advanced Investigator Grant 247333.
\end{acknowledgments}

\end{document}